\documentclass[twocolumn]{article}
\usepackage[margin=2.7cm]{geometry}
\usepackage{graphicx}
\usepackage{amsmath}
\usepackage{cite}
\usepackage{dblfloatfix}
\usepackage{color}
\usepackage{booktabs}

\usepackage{amsmath}
\usepackage{amssymb}
\usepackage{bm}

\title{Deciphering multiple binding modes in large protein complexes via Enhanced Sampling and $\beta$VAE: the CYFIP1–eIF4E case}
\author{Francesco Pignotti$^{1,2}$, Emanuele Telari$^3$, Edoardo Milanetti$^{4,6}$, Anna La Teana$^{1,2}$, Jesmina Rexha$^{1,2}$,
\\ Alice Romagnoli$^{1,2,\ast}$, Luca Maragliano$^{2,5,\ast}$, Giovanni Di Muccio$^{1,2,\ast}$}

\date{}

\begin{document}

\twocolumn[
\begin{center}
{\LARGE \bfseries Deciphering multiple binding modes in large protein complexes via Enhanced Sampling and $\beta$VAE: the CYFIP1--eIF4E case \par}
\vspace{0.4em}
{\normalsize \textit{Running title: CYFIP1--eIF4E binding modes via enhanced sampling} \par}
\vspace{0.6em}
{\large Francesco Pignotti$^{1,2}$, Emanuele Telari$^3$, Edoardo Milanetti$^{4,6}$, Anna La Teana$^{1,2}$, Jesmina Rexha$^{1,2}$,
Alice Romagnoli$^{1,2,\ast}$, Luca Maragliano$^{2,5,\ast}$,
Giovanni Di Muccio$^{1,2,\ast}$ \par}
\vspace{0.8em}
\begin{minipage}{0.9\textwidth}
\centering\footnotesize
$^1$New York-Marche Structural Biology Center (NY-MaSBiC), Polytechnic University of Marche, Via Brecce Bianche, Ancona 60131, Italy\\
$^2$Department of Life and Environmental Sciences, Polytechnic University of Marche, Via Brecce Bianche, Ancona 60131, Italy\\
$^3$Departament de Ci\`encia de Materials i Qu\'imica F\'isica and Institut de Qu\'imica Te\`orica i Computacional, Universitat de Barcelona, Barcelona, Spain\\
$^4$Department of Human Sciences, Link Campus University, Via del Casale di San Pio V 44, 00165 Rome, Italy\\
$^5$Center for Synaptic Neuroscience and Technology, Istituto Italiano di Tecnologia, Genova 16132, Italy\\
$^6$Center for Life Nano \& Neuro Science, Istituto Italiano di Tecnologia, Viale Regina Elena 291, 00161 Rome, Italy\\
$^\ast$Corresponding authors: Alice Romagnoli (a.romagnoli@staff.univpm.it), Luca Maragliano (l.maragliano@staff.univpm.it), Giovanni Di Muccio (g.dimuccio@staff.univpm.it)
\end{minipage}
\end{center}
\vspace{1.5em}
\rule{\textwidth}{0.4pt}
\vspace{0.5em}
\begin{abstract}
Eukaryotic initiation factor 4E (eIF4E) drives cap-dependent translation, regulated by 4E-binding proteins (4E-BPs) competing for its dorsal and lateral surfaces. CYFIP1, a scaffold protein, is both a WAVE Regulatory Complex component and a non-canonical 4E-BP repressing translation via a short helical motif, linked to autism spectrum disorders and cancer. Prior work modeled only this isolated motif as a peptide, leaving the full interface uncharacterized.

Starting from this preliminary evidence, we modeled the full-length proteins as flexible partners: all-atom simulations fed a data-driven docking search (HADDOCK3), which guided funnel-restrained Temperature-Accelerated Molecular Dynamics of a coarse-grained Martini 3 representation. Ten replicas accumulated 100~$\mu$s of enhanced sampling.

Funnel-restraint variables efficiently sample binding and unbinding but are structurally degenerate: direct clustering of the raw trajectory collapses these poses into poorly resolved groups. We instead reweighted the biased trajectories onto a contact-based latent space built by benchmarking PCA against a $\beta$-Variational Autoencoder, then clustered minima of the equilibrium free-energy surface: only intermediate regularization resolved distinct, homogeneous modes.

Three such modes recur, engaging eIF4E's canonical dorsal site, its non-canonical lateral site, and both. We find that CYFIP1 also engages eIF4E through a second surface, the same that mediates its WAVE-complex dimerization, and a basic patch. Both plausibly explain residual binding previously reported after motif mutation. This suggests its translational and cytoskeletal roles compete for the same surface. Most contacts on eIF4E fall outside this dorsal/lateral interface. This heterogeneous, plastic ensemble offers a structural basis for mutagenesis and therapeutic peptide design.
\end{abstract}

\vspace{10pt}

\vspace{10pt}
\textbf{Broader Audience Statement:} Protein synthesis is tightly controlled by regulatory proteins such as CYFIP1, a multifunctional protein whose dysregulation has been linked to autism spectrum disorders and cancer. Using computer simulations and artificial intelligence, we mapped in detail how CYFIP1 grips its translation partner eIF4E, revealing several previously hidden contact regions on both proteins. This structural blueprint could help guide the design of future targeted therapies against this interaction.

\vspace{1em}
\rule{\textwidth}{0.4pt}
\vspace{1.5em}
]

\section{Introduction}

\begin{figure*}[h!]
    \centering
    \includegraphics[width=1\textwidth]{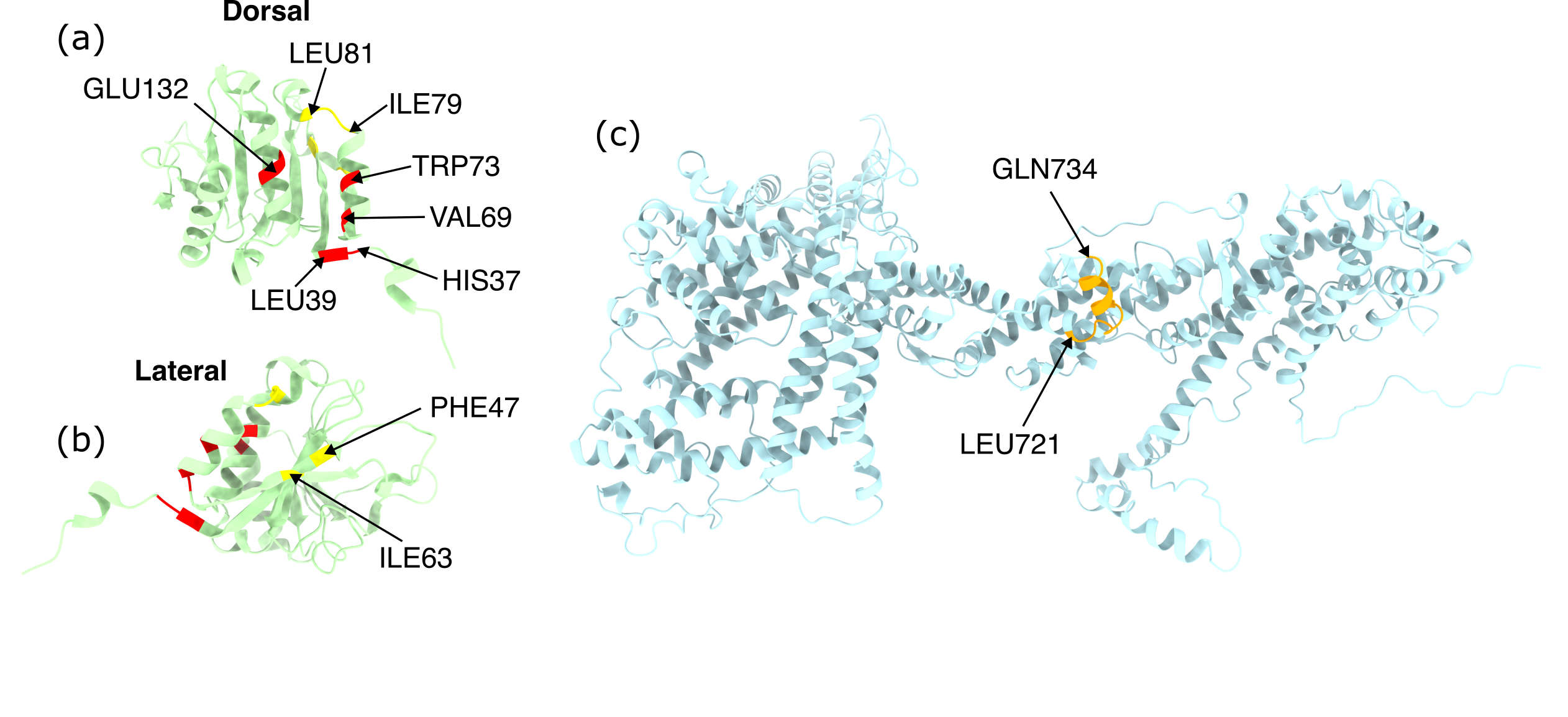}
    \caption{\textbf{eIF4E and CYFIP1 interaction surfaces.} Cartoon representation of the interaction regions between eIF4E (green) and CYFIP1 (cyan), with representative residues of each surface labelled. On eIF4E, the a) dorsal binding surface (e.g. His37, Leu39, Val69, Trp73, Glu132) and b) the lateral surface (e.g. Phe47, Ile63), shared by eIF4G and the 4E-BP family of translational repressors, are highlighted in red and yellow, respectively. c) The short helical eIF4E-binding motif of CYFIP1 (residues 721--734, from Leu721 to Gln734), essential for complex formation, is highlighted in orange.
    }
    \label{fig:intro}
\end{figure*}

Eukaryotic initiation factor 4E (eIF4E) is a cap-binding protein that plays a pivotal role in the regulation of translational initiation. By recognizing the 5$'$ m$^7$GTP cap structure of mRNA, eIF4E recruits the translation machinery binding to eIF4G and promotes ribosome assembly, making it a master regulator of protein synthesis and a key node in the control of cell growth, proliferation, and survival\cite{DiMarino2015}. Its activity is tightly regulated through direct protein-protein interactions with a family of translational repressors known as 4E-binding proteins (4E-BPs). These repressors compete with eIF4G for two binding sites on eIF4E: i) dorsal binding site, Fig.~\ref{fig:intro}a, and ii) lateral surface, Fig.~\ref{fig:intro}b, preventing the assembly of the eIF4F complex and globally suppressing cap-dependent translation \cite{Romagnoli2021}.

CYFIP1 (Cytoplasmic FMR1-Interacting Protein 1) is a multifunctional scaffold protein that operates at the intersection of two major cellular regulatory axes. On one hand, it is a core structural component of the WAVE Regulatory Complex (WRC), a pentameric assembly comprising CYFIP1, WAVE, ABI, NCKAP, and BRK1, which controls actin cytoskeleton remodeling \cite{Han2023}. On the other hand, CYFIP1 was identified as a non-canonical 4E-BP: through a short helical motif located within its sequence (residues 721–734, Fig.~\ref{fig:intro}c), it binds eIF4E and the Fragile X Mental Retardation Protein (FMRP)\cite{Napoli2008}, suppressing cap-dependent translation.

\begin{figure*}[h!]
    \centering
    \includegraphics[width=1\textwidth]{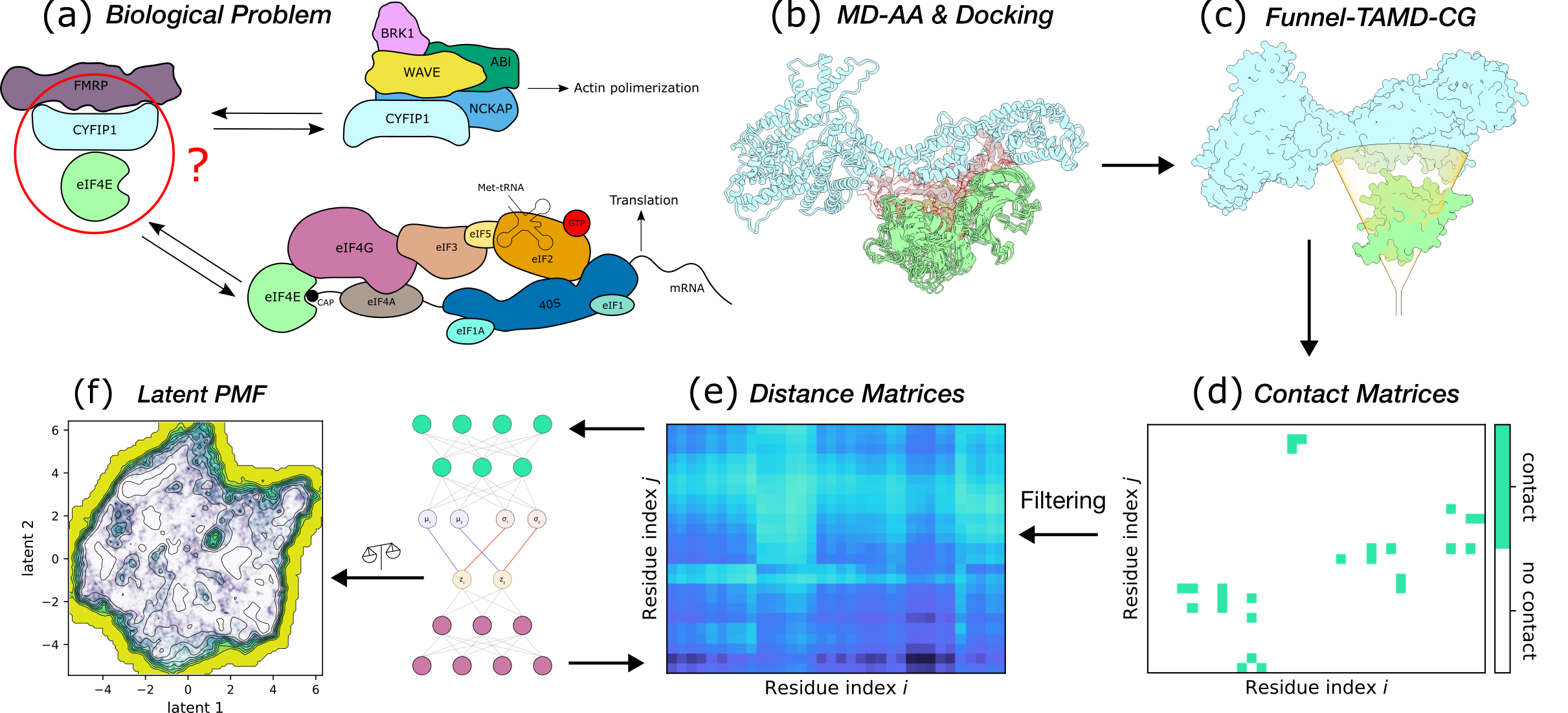}
    \caption{\textbf{Overview of the computational pipeline.} The biological question (a): CYFIP1 acts as a molecular switch between the WAVE Regulatory Complex (WRC; CYFIP1, WAVE, ABI, NCKAP, BRK1), which regulates actin polymerization, and a 4E-BP that sequesters eIF4E and represses cap-dependent translation initiation. (b) Tier 1: all-atom molecular dynamics ensembles of the individual proteins are clustered and used for data-driven docking with HADDOCK3, yielding ensemble models of the CYFIP1 (cyan)--eIF4E (green) complex (interface residues highlighted in red). (c) Tier 2: representative structures are converted to a coarse-grained Martini 3 representation and subjected to funnel-restrained Temperature-Accelerated Molecular Dynamics (TAMD), with the funnel restraint potential centered on the CYFIP1 binding motif; 100 µs of enhanced sampling across 10 replicas allow exploration of the binding/unbinding process and estimation of the binding free energy. Tier 3: (d) binary contact matrices computed from the CG ensemble are filtered to retain the most frequently contacted residue pairs; (e) the corresponding inter-residue distance matrices, restricted to this filtered subset, are used as input features for a $\beta$-Variational Autoencoder ($\beta$-VAE); (f) the resulting two-dimensional latent space, in which the reweighted PMF is used to resolve and extract distinct binding poses.}
    \label{fig:pipeline}
\end{figure*}

This dual functionality, Fig.~\ref{fig:pipeline}a, positions CYFIP1 as a molecular switch  essential for neurogenesis, and synaptogenesis\cite{Chen2010,Han2023}. Its dysregulation has been linked to neurodevelopmental disorders including autism spectrum disorder and schizophrenia\cite{DomnguezIturza2019}.

The structural basis of the CYFIP1-eIF4E interaction remains only partially understood. Mutagenesis studies have established that the helical motif of CYFIP1, spanning residues 721-734, is essential for complex formation; indeed, mutations in this region reduce binding by up to 70\%~\cite{Napoli2008} (Fig.~\ref{fig:intro}c). Previous computational investigations provided an initial structural model of the complex but only considered the helix binding motif as an isolated peptide, without treating the entire CYFIP1 protein~\cite{DiMarino2015,DiMarino2015unique}. Due to the lack of other information, the full extent of the interaction interface and the conformational heterogeneity of the complex have not yet been characterized at atomic resolution. This gap limits both the mechanistic understanding of translational regulation by CYFIP1 and the rational exploitation of this interface for therapeutic purposes.

To address these open questions, we developed and applied a multi-scale computational pipeline specifically designed to handle the conformational flexibility of a large protein-protein complex, a challenge shared with other large, flexible multidrug transport systems whose substrate binding has been characterized through molecular dynamics simulations \cite{Vargiu2012}. The strategy is organized in three sequential tiers. In the first tier, extensive all-atom molecular dynamics simulations of the individual proteins were used to generate different structural ensembles, which were subsequently clustered and used for a data-driven docking protocol using HADDOCK3 \cite{Giulini2025}. The resulting ensemble complexes served for defining the geometrical parameters for a specific funnel-shaped spatial restraint potential around the known binding motif of CYFIP1 (Fig.~\ref{fig:pipeline}b) used in the second tier. The most representative docked structures were converted to a coarse-grained representation using the Martini 3 force field \cite{Souza2021} and subjected to Temperature-Accelerated Molecular Dynamics (TAMD)\cite{Maragliano2006} in combination with a funnel restrain~\cite{Limongelli2013}, accumulating 100~$\mu$s of enhanced sampling across 10 replicas, Fig.~\ref{fig:pipeline}c. This approach enables multiple binding and unbinding events, thereby ensuring extensive sampling of the binding free energy landscape, in line with enhanced-sampling strategies developed for resolving protein-protein binding pathways and rate constants \cite{Saglam2019}. In the third tier, the configurational ensemble generated by TAMD simulations was first reduced to binary contact matrices of the most frequently contacted residues (Fig.~\ref{fig:pipeline}d), converted into inter-residue distance matrices (Fig.~\ref{fig:pipeline}e), and analyzed using a combination of $\beta$-Variational Autoencoder ($\beta$-VAE) and free-energy reweighting, addressing the broader challenge of extracting structure from the high-dimensional, time-correlated data generated by molecular dynamics simulations through deep learning \cite{Bonneau2026}.
%considering only the most contacted residue-residues pairs estimated with a contacts matrices \ref{pipeline}d, with the aid of mlcolvar \cite{Bonati2023} library.

By systematically comparing multiple dimensionality reduction strategies (including PCA, a standard autoencoder, and a range of $\beta$-VAE models), we identified an optimal latent representation for the complex, Fig.~\ref{fig:pipeline}f, enabling the resolution of distinct and structurally homogeneous binding poses embedded within an otherwise degenerate configurational space. Together, this pipeline provides an unprecedented picture of the conformational and energetic landscape of the CYFIP1-eIF4E complex, identifies possible previously uncharacterized interaction regions on both partners, and lays the quantitative groundwork for future mutagenesis experiment.

\section{Results}

%%%%%%%%%%%%%%%%%%%%%%%%%%%%%%%%%%

\subsection{Sampling convergence and PMF}
\begin{figure*}[h!]
    \centering
    \includegraphics[width=1\textwidth]{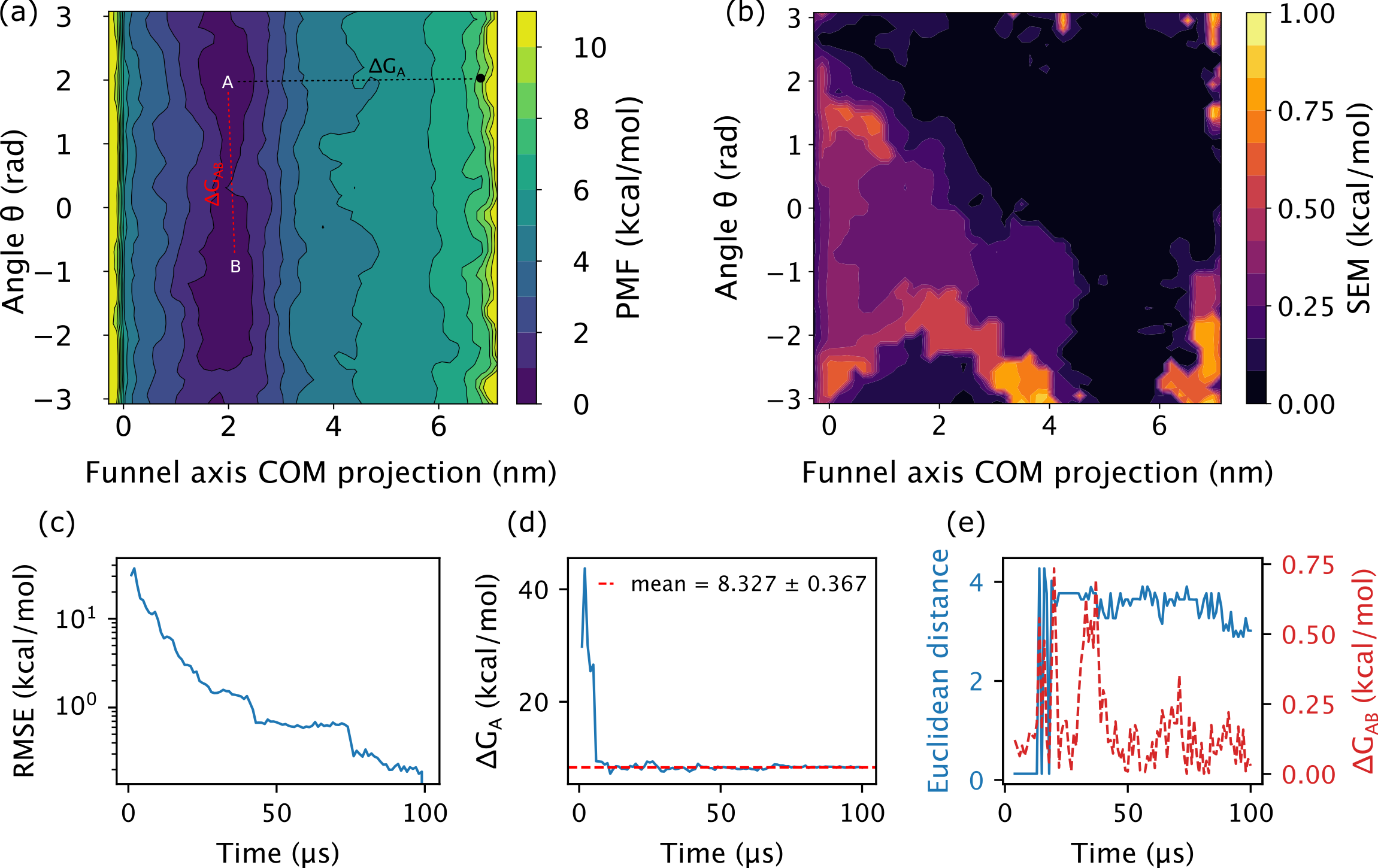}
    \caption{\textbf{Free-energy landscape of the CYFIP1--eIF4E complex and its convergence.} (a) Potential of mean force (PMF) reconstructed from the funnel-restrained TAMD simulations, projected onto the two collective variables: the center-of-mass displacement of eIF4E along the funnel axis ($x$) and the torsional angle $\theta$ defined between CYFIP1 and eIF4E ($y$); energy is given in kcal/mol (color bar). The two bound-state minima, A and B, are indicated, together with the free-energy differences $\Delta G_A$ (black dotted line, from minimum A to the unbound reference point) and $\Delta G_{AB}$ (red dashed line, between minima A and B), quantified as a function of simulation time in panels (d) and (e), respectively. (b) Associated standard error of the mean (SEM) over the same CV space. (c) Time evolution of the root-mean-square error (RMSE) of the PMF in panel (a), computed with respect to the PMF from the last simulation frame, as a function of cumulative simulation time. (d) Time evolution of the depth of the free-energy well at minimum A, $|\Delta G_A|$, estimated as the PMF difference between the unbound reference point, located at $(6.832, 2.163)$ in CV space and representing the unbound state in solution, and minimum A; plotted as a positive quantity, it corresponds to the binding free energy reported in the main text with the standard sign convention, $\Delta G_A = -8.327 \pm 0.367$ kcal/mol. The mean value and its associated error were computed as for the PMF, relative to the last 20~$\mu$s of simulation. (e) Time evolution of the Euclidean distance in CV space (blue line, left axis) and of the free-energy difference $\Delta G_{AB}$ (red dashed line, right axis, kcal/mol) between minima A and B.}
    \label{fig:fe_convergence}
\end{figure*}

TAMD simulations reconstructed the CYFIP1–eIF4E binding PMF landscape as a function of two collective variables: i) the center-of-mass displacement of eIF4E with respect to CYFIP1 projected along the funnel axis, and ii) the torsional angle $\theta$ defined by two points on CYFIP1 and two points on eIF4E, see Methods and Fig. S1 for details. The total sampling time is $100\,\mu$s computed via 10 independent replicas of 10 $\mu$s.

The resulting landscape is reported in Fig.~\ref{fig:fe_convergence}a, and 
the standard error computed over the last $20\,\mu$s of samplig is reported in Fig.~\ref{fig:fe_convergence}b, showing that the maximum error is lower than 1 kcal/mol in almost all the landscape.

The bound-state basin is located at a CYFIP1-eIF4E distance of about 2~nm and contains the main energy minimum, within which 
two states can be identified: state A ($\theta \approx +2$~rad) and state B ($\theta \approx -1$~rad). 
The two minima lie within the same elongated low energy lobe, and are separated by less than 1 kcal/mol rather than by a large free energy barrier, indicating that the poses can easily move 
between each other. 
The bright yellow stripe at distance 0~nm corresponds to a hard repulsive wall, consistent with the bottom of the funnel and uniform along the angular coordinate.

Moving outward, the PMF loses the angular dependence, developing angle-isotropic plateau, as expected. Indeed, once the protein leaves the specific contacts of the binding site, the rotational orientation no longer modulates the energetics and the torsional angle effectively becomes a non-discriminating collective variable. The yellow stripe at $x \approx 7$~nm does not represent a physical barrier, but rather the artificial confinement imposed by the funnel restraint potential.
No appreciable unbinding barrier is present along the funnel axis: the surface rises smoothly and monotonically from the bound minimum to the bulk plateau.
The PMF convergence was monitored computing the root-mean-square error (RMSE) with respect to PMF computed at the last frame, reaching a value lower than 0.2 kcal/mol in the last 20 $\mu$s, see Fig.~\ref{fig:fe_convergence}c. Also monitoring the PMF difference between the bound and unbound states, and between the two minima (A and B), shows that the PMF converged after approximately 30 $\mu$s of sampling, Fig.~\ref{fig:fe_convergence}d-e. The estimated binding free energy for the global minimum A is $\Delta G_A = -8.327 \pm 0.367\,\mathrm{kcal/mol}$ (Fig.~\ref{fig:fe_convergence}d); the free-energy difference between minima A and B, $\Delta G_{AB}$, remains small (well below 0.3 kcal/mol) over the last 20 $\mu$s of sampling (Fig.~\ref{fig:fe_convergence}e), consistent with the elongated, weakly partitioned low-energy region described above.

%%%%%%%%%%%%%%%%%%%%%%%%%%%%%%%%%%%%%%%%%%
\subsection{Contact Analysis}
\begin{figure*}[h!]
    \centering
    \includegraphics[width=1\textwidth]{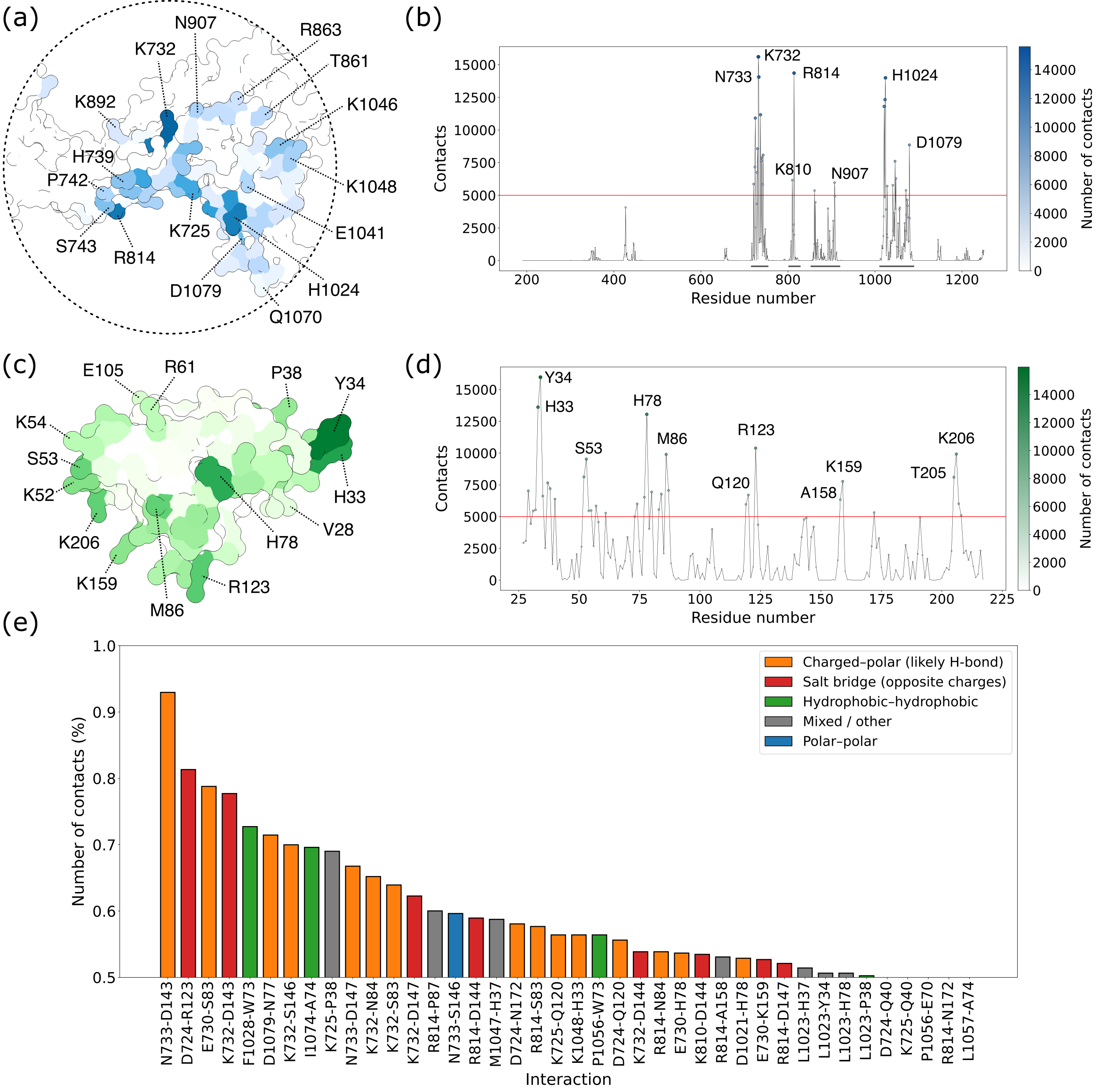}
\caption{\textbf{Contact analysis of the CYFIP1--eIF4E complex.} Analysis performed along the funnel-restrained TAMD ensemble. (a) Surface representation of CYFIP1 colored by per-residue contact frequency (white to blue, low to high). (b) Per-residue contact counts for CYFIP1; each point is a residue and the red line marks the threshold above which residues are considered strongly involved in the interface. (c,d) Same representations for eIF4E (white to green surface, and per-residue contact distribution with threshold). (e) Most recurrent residue--residue contacts (CYFIP1--eIF4E), expressed as the fraction of frames in which the contact is formed (\%); bars are colored by the chemical nature of the interaction: charged--polar (likely H-bond, orange), salt bridge (red), hydrophobic--hydrophobic (green), polar--polar (blue), and mixed/other (gray).}
    \label{fig:interactions}
\end{figure*}

Given the large amount of data generated ($102,305$~frames), the first step in our analysis was to calculate a binary contact matrix for each simulation frame. This approach allows us to identify the residues most critically involved in complex formation.

Focusing on CYFIP1, different regions exhibit strong signals, as shown in Fig.~\ref{fig:interactions}a,b. The strongest signal spans residues 721--734. This region corresponds to a previously, experimentally characterized helix motif of CYFIP1 that was shown to bind eIF4E. Indeed, using the pull-down technique, it was shown that mutations within this motif reduce complex formation by up to 70\% \cite{Napoli2008}.

A second spot shows two other potentially important basic residues, Lys810 and Arg814, that are part of an extended helix below the eIF4E binding motif; in particular, residue Arg814 is the second most contacted residue after Lys732.

A third notable region extends from residues 1010--1090, which we designate the CYFIP1 Dimerization/Interaction Module (CDIM) domain in the following text. This region comprises a larger structural segment of CYFIP1, including multiple helices and a loop region. Interestingly, this portion of the protein is typically involved in CYFIP1/2 dimerization, a critical step for WRC assembly~\cite{Chen2010}. Here, His1024 is present as the third most contacted residue, approximately tied with Asn733, in our sampling.

A further region displaying moderate contact frequency involves Asn907 together with the neighboring residues Thr861 and Arg863 (Fig.~\ref{fig:interactions}a,b), located at the edge of the interaction surface, yet still contiguous with the main cluster of contacts, between the helix motif and the CDIM domain. Although less frequently contacted than these two regions, this marginal patch extends the CYFIP1--eIF4E interaction surface beyond its core, suggesting a minor but non-negligible role in complex formation.

With regard to eIF4E (Fig.~\ref{fig:interactions}c,d), several residues form a high number of contacts with CYFIP1, including His33, Tyr34, His78, and Arg123. These residues are of particular interest, as they belong to both the canonical and non-canonical binding sites of 4E-binding proteins (highlighted bands). This evidence, combined with the observation that these interactions occur with the helix and CDIM regions, further supports our results.

We also analyzed residue-specific contacts to identify the most relevant residue--residue interactions occurring during the simulations. As shown in Fig.~\ref{fig:interactions}e, the strongest and most recurrent interactions involve the helix motif and the CDIM of CYFIP1, consistent with the per-residue contact counts reported in Fig.~\ref{fig:interactions}b,d. In particular, residue eIF4E:Asp143 is especially noteworthy, as it is involved in the first and fourth most frequent interactions. Residues CYFIP1:Lys724, Lys725, Lys732, and Asn733, belonging to the helix motif, participate in many of the most significant interactions. Within the CYFIP1 CDIM domain, residues Phe1028 and Asp1079 display the two most recurrent interactions. Notably, although eIF4E:Tyr34 shows the highest overall contact count in the general per-residue analysis (Fig.~\ref{fig:interactions}d), it does not appear among the strongest pairwise interactions in Fig.~\ref{fig:interactions}e: its most frequent partner, CYFIP1:Leu1023, ranks toward the lower end of the distribution. This indicates that eIF4E:Tyr34 engages in a broad, distributed pattern of contacts across multiple CYFIP1 residues rather than a single dominant interaction.

%%%%%%%%%%%%%%%%%%%%%%%%%%%%%%%%%%%%%%%%%%%%%%%%%%%%%%
\begin{figure*}[hbtp!]
    \centering
    \includegraphics[width=0.9\textwidth]{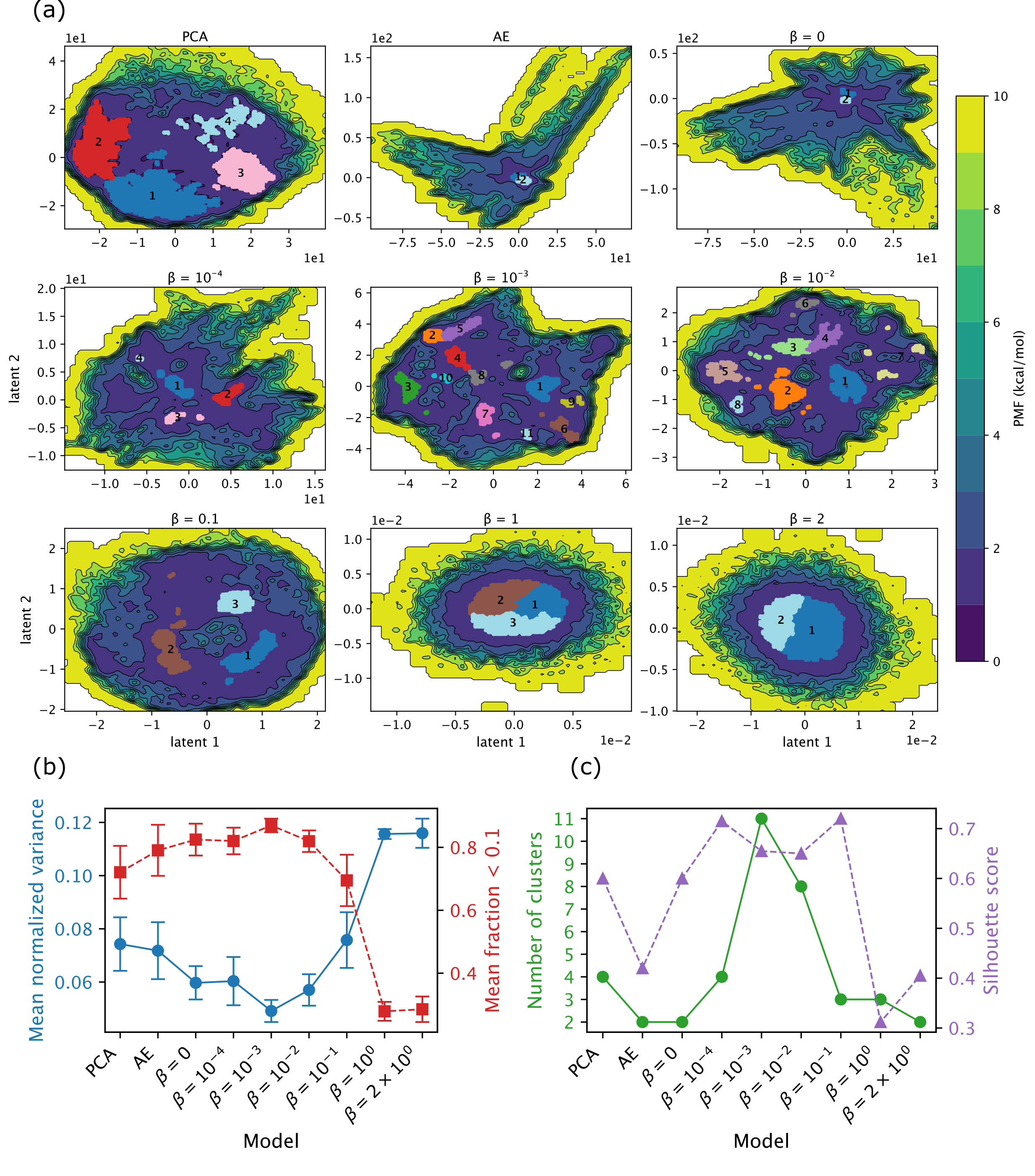}
\caption{\textbf{Clustering of the reweighted latent space.} Comparison of dimensionality reduction models for resolving the
binding-pose degeneracy: PCA, a standard autoencoder (AE), and $\beta$-VAE
models with $\beta = 0$, $10^{-4}$, $10^{-3}$, $10^{-2}$, $0.1$, $1$ and $2$.
(a) Reweighted PMF (kcal/mol, see colour bar) in the two-dimensional latent
space of each model; coloured patches mark the clusters of the low-free-energy
region, numbered by decreasing population. 
Note the model-dependent latent scale on each axis, spanning orders of magnitude from $10^{-2}$ (e.g. $\beta=1,2$) to $10^{2}$ (AE, $\beta=0$).
(b) Mean normalized variance 
$\langle \widetilde{\sigma_{ij}^2} \rangle$ from Eq.~\eqref{eq:mean_variance} (blue line, lower is better)
and mean fraction $\langle \theta \rangle$ of
$\widetilde{\sigma_{ij}^2}<0.1$, Eq.~\eqref{eq:mean_fraction}
(red line, higher is better); 
error bars are the standard error of the mean across clusters.
(c) Number of clusters (green, left axis) and best silhouette score 
(purple, right axis) computed in each model's own latent space, thus measuring geometrical separation within that representation.
}
\label{fig:rew_pmf}
\end{figure*}

%%%%%%%%%%%%%%%%%%%%%%%%%%%%%%%%%%%%%%%%%%%%%%%%%%%%%%%%
\subsection{Resolving Binding Mode Degeneracy: Latent-Space Potential of Mean Force Clustering}

The convergence analysis of the free-energy surface (Fig.~\ref{fig:fe_convergence}a) established that TAMD sampling reliably resolves two dominant minima along the chosen collective variables. The contact analysis (Fig.~\ref{fig:interactions}b,d) further identified the residue-residue interactions most frequently populated during the dynamics, pointing to several candidate binding regions on both CYFIP1 and eIF4E. Taken together, however, these analyses do not fully characterize the conformational ensemble of the complex. Each free-energy minimum is not a single, structurally homogeneous state, but an envelope of multiple distinct relative orientations of CYFIP1 and eIF4E: the chosen CVs, while adequate to capture the overall binding/unbinding coordinate, are degenerate with respect to the internal orientational degrees of freedom of the complex. Consistently, different poses can share a subset of contacts while differing substantially in others, so that contact frequency alone cannot disambiguate them.

Disentangling these overlapping conformations is therefore non-trivial. A \textsl{naive} clustering performed directly on the configurational dataset -- using either the raw distance matrices or the most relevant contact frequencies as input -- yields low clustering performance: the best silhouette score is 0.16, obtained for three clusters, all showing low structural homogeneity (Fig.~S2). Neither the free-energy surface nor the contact map, individually, thus provides a representation capable of separating the distinct binding modes populating the ensemble.

To address this, we adopted a two-step strategy combining dimensionality reduction with free-energy reweighting, building on the general strategy of using nonlinear dimensionality reduction to resolve free-energy landscapes from molecular dynamics data \cite{Das2006}. First,
we trained a $\beta$-Variational Autoencoder ($\beta$-VAE),
similar to the protocol proposed by \cite{adhikari2025elucidating}, on distance matrices from the most frequently contacting residues, re-projecting the high-dimensional configurational space into a two-dimensional collective variable (CV) space capable of separating and describing the different conformations. Projection alone was not sufficient, however: clustering directly on the latent space remained non-trivial due to numerous intermediate states (Fig.~\ref{fig:pipeline}f). We therefore reweighted the TAMD trajectory to reconstruct the equilibrium PMF as a function of the latent variables, and clustered the low-free-energy regions of this reweighted surface -- rather than the raw latent projections -- to extract the structures corresponding to the minima and isolate a set of distinct, well-resolved conformations.

To identify the optimal model for this task, we compared multiple dimensionality reduction approaches: standard PCA, a standard autoencoder (AE), and a series of $\beta$-VAE models spanning regularization strengths $\beta = 2, 1, 10^{-1}, 10^{-2}, 10^{-3}, 10^{-4}$, and $0$, reconstructing the corresponding 2D PMF surfaces in the latent space (Fig.~\ref{fig:rew_pmf}a). For each model, cluster quality was assessed with two structural metrics: i) the mean normalized variance $\langle \widetilde{\sigma^2_{ij}} \rangle$ of the distance matrix of each cluster, reflecting overall cluster homogeneity, and ii) the mean fraction $\langle \theta \rangle$ of $\widetilde{\sigma^2_{ij}}$ values below 0.1, both reflecting intra-cluster structural consistency (Fig.~\ref{fig:rew_pmf}b), together with the number of resolved clusters and the corresponding best silhouette score (Fig.~\ref{fig:rew_pmf}c).

The $\beta$-VAE with $\beta = 10^{-3}$ is the best-performing model, jointly minimizing $\langle \widetilde{\sigma^2_{ij}} \rangle$ and maximizing $\langle \theta \rangle$ (Fig.~\ref{fig:rew_pmf}b). This optimum coincides with a peak in the number of resolved clusters, reaching 11 against 2--8 for all other models (Fig.~\ref{fig:rew_pmf}c), together with a high silhouette score (0.65), indicating that the additional clusters are genuinely well separated rather than a by-product of over-partitioning. The silhouette score itself peaks slightly higher at $\beta = 10^{-4}$ and $\beta = 10^{-1}$ (0.71), reflecting the trade-off between resolving more numerous, individually more compact clusters and the coarser partitions of neighboring models.

Clustering performance degrades gradually, not abruptly, around this optimum. Moving away from $\beta = 10^{-3}$ in either direction, both structural metrics remain close to their optimal values over a broad range ($10^{-4}$ to $10^{-2}$; Fig.~\ref{fig:rew_pmf}b), and the corresponding PMF surfaces retain multiple, well-resolved low-free-energy regions (Fig.~\ref{fig:rew_pmf}a). The number of resolved clusters is more sensitive to $\beta$, dropping to 4 and 8 at $\beta = 10^{-4}$ and $\beta = 10^{-2}$, respectively, though still with good silhouette scores and visibly structured latent spaces. $\beta = 10^{-3}$ thus marks a genuine sweet spot, not an isolated point within an otherwise poorly performing neighborhood.

A markedly more severe degradation occurs at the extremes of the $\beta$ range, through two distinct mechanisms. For $\beta = 2$ and $\beta = 1$, both structural metrics deteriorate sharply, and the optimal number of clusters collapses to one, indicating that the latent space is as degenerate as the original CV space, with no natural separation into distinct states. Forcing a partition into 2--3 clusters still yields poor silhouette scores (0.31--0.38), confirming that this lack of structure is genuine rather than an artifact of cluster-number selection. The latent coordinates span a remarkably narrow range (on the order of $10^{-2}$), and the resulting PMFs exhibit a single, nearly Gaussian basin centered at the origin -- the signature of an over-regularized regime in which the Kullback-Leibler term dominates the loss function, collapsing the encoder distribution onto the standard normal prior and discarding most of the discriminative information in the input distance matrices.

At the opposite extreme, the standard AE and the $\beta = 0$ model both fail to resolve genuine binding modes, though not exactly to the same extent. Structural metrics are only moderately worse than at the optimum for both, and in both cases the latent space collapses into a single, elongated, filamentous low-free-energy region devoid of internal free-energy barriers, rather than a compact basin. As for $\beta = 2, 1$, the best partition in both cases is a single cluster; forcing a division into 2 clusters yields a markedly better silhouette score for $\beta = 0$ (0.60) than for the AE (0.40), suggesting that some residual structure survives in the $\beta$-VAE even at vanishing regularization, likely due to the stochastic bottleneck retained in the architecture. The overall shape and scale of the two PMFs remain qualitatively similar, consistent with the theoretical expectation that, as $\beta \to 0$, the regularization term vanishes from the objective and the reconstruction loss alone favors a deterministic encoding, formally reducing the $\beta$-VAE objective to that of a standard AE, though this reduction is only exact in the limit, and small quantitative differences persist in practice.

The PMF derived from PCA displays intermediate behavior: a few local minima can be identified, arranged along nearly rectilinear directions, with markedly less separation than in the $\beta = 10^{-3}$ case. This reflects the intrinsic limitation of PCA as a linear technique, which captures only the directions of maximal variance and cannot disentangle the non-linear manifold along which the conformational ensemble is distributed, indicating that the conformational heterogeneity sampled along the TAMD trajectories is genuinely non-linear, and that an appropriately regularized probabilistic latent representation is required to resolve the distinct binding poses of the complex.

%%%%%%%%%%%%%%%%%%%%%%%%%%%%%
\begin{figure*}[h!]
    \centering
    \includegraphics[width=1\textwidth]{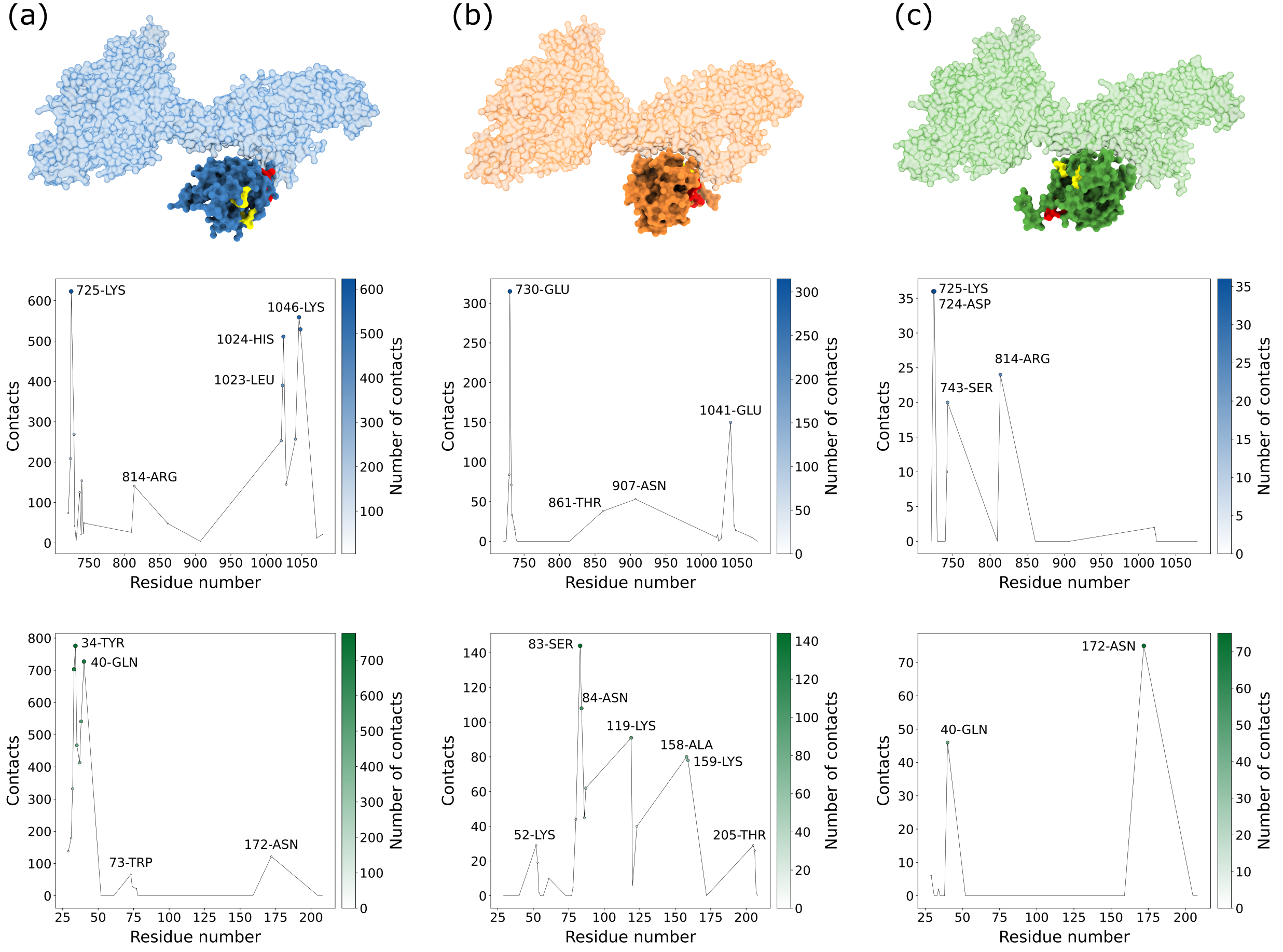}
\caption{\textbf{Representative binding poses of the three most populated clusters.} Structures identified in the $\beta = 10^{-3}$ latent space, for the CYFIP1--eIF4E complex. Each column shows the pose displaying the largest number of contacts within its cluster: (a) Cluster~1 (blue), (b) Cluster~2 (orange), and (c) Cluster~3 (green). (Top) Surface representation of the complex; on eIF4E the canonical 4E-BP binding site (dorsal surface) is highlighted in red and the non-canonical binding site (right lateral surface) in yellow. (Middle) Per-residue contact counts of CYFIP1
and (bottom) of eIF4E, for each cluster, with the most contacted residues labelled; color intensity encodes the number of contacts. The interaction is dominated by the dorsal surface of eIF4E in Cluster~1 (a), shifts to the right lateral surface in Cluster~2 (b), and additionally engages the outer ventral region (outside the m$^{7}$GTP cap-binding pocket) in Cluster~3 (c), while a residue of the CYFIP1 helical motif acts as a contact hotspot in all three clusters.}
    \label{fig:clusters}
\end{figure*}

\subsection{Possible binding mode picture}
Having identified the $\beta$-VAE latent space with $\beta = 10^{-3}$ as the optimal representation, 
all frames residing within each energy minimum were extracted from the corresponding structural 
ensemble and subjected to a binary contact analysis. 

For each cluster, the complex structure displaying the largest number of contacts is shown in Fig.~\ref{fig:clusters}.
To aid the interpretation of the interaction patterns, we define the spatial anatomy of eIF4E as follows: the CAP-binding domain constitutes its ventral surface, the canonical 4E-BP binding  site its dorsal surface (shown in red in Fig.~\ref{fig:intro} and Fig.~\ref{fig:clusters}), and the non-canonical 4E-BP binding site its right lateral surface (shown in yellow in Fig.~\ref{fig:intro} and Fig.~\ref{fig:clusters}).

Turning first to the residues of CYFIP1 that contact eIF4E, all three most populated clusters (Fig.~\ref{fig:clusters}) share at least one residue from the helical motif of CYFIP1 as a 
high-interaction hotspot. The region spanning residues 1010-1090 consistently emerges as a contact-rich area in the all-trajectory contact analysis and in this latter re-emerge with most notably contacts through \textsc{Leu1023}, \textsc{His1024}, and 
\textsc{Lys1046} in Cluster1 (a). \textsc{Arg814} recurs as a contact in both Cluster1 and Cluster~3 (c), whereas \textsc{Thr861} and \textsc{Asn907} appear exclusively in Cluster2 (b).
The recurrence of these contacts is particularly noteworthy given that this analysis incorporates energetic weighting, and thus implicitly reflects the statistical likelihood of each pose.

Turning to the residues of eIF4E that contact CYFIP1, and as further illustrated by the structures shown in Fig.~\ref{fig:clusters}, the residues classically described as 
the 4E-BP interaction interface are not engaged. The sole exception is \textsc{Trp73} a canonical 4E-BP contact documented in the literature which appears in Cluster1. 
Nevertheless, a coherent interaction pattern emerges when the broader structural anatomy of eIF4E is considered. In Cluster1, the primary interaction involves the dorsal surface of eIF4E, a finding consistent with the contact analysis performed over the full trajectory 
and highlighted by \textsc{Tyr34} in Cluster1 and \textsc{Gln40} in both Cluster~1 and Cluster~3. In Cluster~2, the dominant interaction shifts to the right lateral surface. In Cluster~3, eIF4E additionally shows partial engagement with the outer ventral region, 
outside the m$^7$GTP cap-binding pocket, most prominently through \textsc{Asn172}.

\section{Discussion and Conclusion}
The results presented in this work provide a comprehensive structural characterization of the CYFIP1–eIF4E complex. The multi-scale pipeline allowed us to treat both partners as semi-flexible proteins and to explore the multiple binding/unbinding processes. 

The recurrence of the helical motif as a primary binding hotspot, in the 100~$\mu$s extended MD sampling, provides a structural rationalization of the experimental observation that mutations within residues 721–734 reduce complex formation by up to 70\%, while the residual binding capacity may now be attributed to the auxiliary surfaces identified in this study, in particular the 1010–1090 region and the basic patch comprising LYS810 and ARG814.

The fact that the 1010–1090 segment also mediates CYFIP1/2 dimerization within the WRC, as can be observed in the crystal structure (PDB: 3P8C) \cite{Chen2010}, suggests a potential mechanistic coupling between the translational and cytoskeletal functions of CYFIP1, with eIF4E and the WRC partners possibly competing for partially overlapping surfaces.

On the eIF4E side, the interaction pattern further reinforces the view of CYFIP1 as a non-canonical 4E-BP. With the notable exception of TRP73, the residues classically described as the 4E-BP binding interface are largely not engaged, and eIF4E instead interacts with CYFIP1 through its dorsal, right-lateral, and outer ventral surfaces in a cluster-dependent fashion, with residues such as TYR34, GLN40, ASP143, and ASN172 emerging as recurrent contact points.

From a methodological perspective, the systematic comparison of dimensionality reduction strategies indicates that the conformational manifold of the complex is genuinely non-linear and that a probabilistic, mildly regularized latent representation is necessary to disentangle it. The monotonic trend observed as a function of $\beta$, from over-regularized Gaussian collapse at $\beta 2$ to unstructured filamentary distributions at $\beta = 0$, provides a practical criterion for selecting the regularization strength in analogous problems, as demonstrated also in other works. 

% Notably, the optimal latent space proved sensitive to orientational differences but \tr{comparatively insensitive to translational displacements} within the bound basin, in line with the physical structure of the free energy surface and suggesting that the $\beta$-VAE has effectively learned a representation that captures the chemically meaningful degrees of freedom of the interaction.

A central outcome of this study is the redefinition of the CYFIP1–eIF4E interface as an intrinsically heterogeneous and plastic ensemble of binding modes, rather than a single well-defined pose. 

Several limitations should nevertheless be acknowledged. The coarse-grained Martini 3 representation, while essential to access the required time scales, smooths out side-chain-specific contributions, and the elastic network restricts large-scale conformational rearrangements and potential allosteric responses. The binding affinities should therefore be considered, as usual in simulations, as relative rather than absolute reference values. Finally, the choice of input features for the $\beta$-VAE introduces an unavoidable bias toward the regions of the interface that are already well sampled, and rarely visited binding modes may remain underrepresented.
Despite these limitations, the picture that emerges has clear implications for both fundamental biology and translational research. The identification of multiple, energetically comparable binding modes and previously uncharacterized contact regions provides a rational basis for the design of mutagenesis experiments and offers concrete structural templates for the optimization of CYFIP1-derived peptide, which have recently been proposed as therapeutic tools against cancer and Fragile X Syndrome \cite{Romagnoli2024,Romagnoli2021peptides,Romagnoli2025}. More broadly, the pipeline introduced here is fully transferable to other flexible protein–protein systems in which multiple low-energy binding modes are expected to coexist.

\section{Materials and Methods}

All simulations were carried out using Gromacs 2022.3 \cite{Abraham2015} patched with PLUMED 2.8 \cite{Tribello2014} compiled with DRR \cite{Chen2018} and funnel plugins\cite{Limongelli2013,Raniolo2020}.
All the plots and analysis were generated using python scripts. 

\subsection{All-atoms simulations} 

The two PDB used to starting the simulations were, respectively, 1WKW \cite{Tomoo2005} for eIF4E and 3P8C \cite{Chen2010} for CYFIP1. The missing loops of CYFIP1 were modeled using the SWISS-MODEL web server \cite{Waterhouse2018}. The system preparation was carried out using AmberTools24 \cite{Case2023}. Residue numbering for CYFIP1 throughout this work follows the PDB 3P8C / UniProt Q7L576 sequence. This numbering differs from that used in the original biochemical characterization of the CYFIP1 eIF4E-binding motif \cite{Napoli2008}, which is based on an earlier UniProt entry (Q14467); residue correspondence between the two schemes was confirmed by direct sequence alignment, e.g. the mutated peptide AGSLLLDKRLRSECKNQ used in \cite{Napoli2008} corresponds to residues 718--734 in the numbering adopted here.

Initially, two different all-atoms systems were generated, one for eIF4E and one for CYFIP1. The force field used was Amber ff15ipq \cite{Debiec2016}. The systems were solvated using the SPC/E water model \cite{Takemura2012} and neutralized at a concentration of 150 mM of NaCl. 
 
The equilibration protocol for all-atoms simulations consists of a first minimization using the steepest descent method that reaches convergence after 6025 steps. 

Three different simulations were carried out to stabilize the systems, consisting of 1 step in the NVT ensemble and 2 steps in the NPT, for a total time of 1 ns of equilibration. An initial restraint of 1000 kj/mol was applied in the first two steps to the proteins backbone and decreased to 0 kj/mol in the final step. The leap-frog algorithm was used for integration, and the v-rescale thermostat and c-rescale barostat were employed for temperature and pressure control at 300K and 1 bar for both equilibration and production. This latter consists of 3 replicas in NVT ensemble of 1 $\mu$s, for a total of 3 $\mu$s of time. 
The cutoff-scheme used was the Verlet list with a neighbor list update every 200 fs (100 steps). Van der Waals interactions were calculated using a cut-off of 1.0 nm. Electrostatic interactions were treated using the particle-mesh Ewald (PME) method with a real-space cutoff of 1.0 nm. All covalent bonds involving hydrogen atoms were constrained using the LINCS algorithm. 
 
Clustering of the simulations was performed using hierarchical clustering. The linkage matrix was calculated based on the RMSD of the backbone atoms, employing the Ward method, resulting in the identification of seven distinct clusters for both CYFIP1 and eIF4E. 

\subsection{Docking procedure}

The docking procedure was performed using HADDOCK3 \cite{Giulini2025}, employing representative structures of CYFIP1 and eIF4E obtained from clustering analyses. A total of 49 docking runs were carried out, corresponding to all possible pairwise combinations of the cluster medoids. Residues 724–732 of CYFIP1 and residues 47 and 106 of eIF4E were defined as active for the docking procedure, based on previous in vitro and in silico studies that highlighted their essential role in the interaction \cite{Napoli2008,DiMarino2015}.

In each docking run, a defined set of HADDOCK3 modules was employed. First, a rigid-body docking stage was performed, with the maximum sampling number set to 200 and the tolerance value set to 5. The {\sl randremoval} option was enabled. The resulting poses were then clustered using the {\sl clustfcc} module and filtered with {\sl seletopclusts}, retaining only the best five poses from the top ten clusters. All selected poses were subsequently subjected to {\sl flexref} and {\sl emref} refinement procedures.

All the generated complexes, for a total of 699 structures, were used to calculate the mean center of mass (COM) of the interaction surface of CYFIP1 and eIF4E. The atoms at the interface are the ones having a distance lower than 0.6 nm between the two proteins, shown as the red surface in Fig.~S1a. The procedure yielded two distinct points: one corresponding to the COM of the CYFIP1 interface side, and the other to the COM of the eIF4E interface side (Fig.~S1b, Tab.~S1).

\subsection{Coarse-graining procedure}

CYFIP1 and eIF4E structures were converted to coarse-grained representations using the {\sl martinize2} script \cite{Kroon2025}. 
The Martini 3 force field~\cite{Souza2021} was used for all simulations. Each system was solvated using the {\sl insane3} script in a 150 mM NaCl solution and equilibrated following an adapted version of the protocol described in~\cite{DiMarino2023}. The time step was set at 20 fs.

To identify the optimal elastic network (EN), five different configurations were generated for both systems, and a 1 µs production simulation was performed for each of them. To account for its intrinsic flexibility, the inter-interaction of the CYFIP1 loop region that span to residues 1017 from 1093, was treated using a modified EN, with a fix potential of 200 $\mathrm{kJ/mol/nm^2}$, see Fig.~S3.

The best-performing model was selected based on its ability to reproduce the RMSF (Fig.~S3a-b) and RMSD (Fig.~S3c-d) profiles obtained from all-atom simulations. The best model is defined by the following parameters: lower bound = 0 nm, decay factor = 1, decay power = 1, upper bound = 0.9 nm, and an elastic bond force constant of 700 $\mathrm{kJ/mol/nm^2}$, except for the the inter-interaction loop CYFIP1 as described above.

\subsection{Coarse-grained funnel TAMD setup}  

Temperature accelerated molecular dynamics (TAMD)\cite{Maragliano2006} was used to enhance sampling of binding/unbinding transitions. To focus TAMD sampling on the region detected by the docking procedure, a funnel shaped potential was applied, see Fig.~S1d. Two collective variables (CVs) were used: (i) the projection of the eIF4E COM along the funnel axis, and (ii) a dihedral angle (Fig.~S1c) defined using four reference point, corresponding to the COM of eIF4E, the COM of a group of atoms within eIF4E, and the COMs of two groups of CYFIP1 atoms, reported in Tab.~S1. The funnel axis was defined using the two reference points in space, calculated as described in the previous docking section. The point dividing the cylindrical and the conical volumes, the cone’s aperture, the lower and upper bound, and the cylinder radius were reported in Tab.~S2. These values were chosen to ensure that the conical region encompasses the entire radius of gyration of eIF4E, which is approximately 7.6 nm. This configuration allowed eIF4E to adequately sample both the solvent environment and the interaction surface. In TAMD, spring forces of 24 kcal/(mol$\cdot$nm$^2$) between the CVs and the fictitious variables were used, with an effective friction of 5000 ps$^{-1}$ and an effective temperature of 5200 K, corresponding to a barrier of approximately 10.34 kcal/mol. This latter value was selected based on the time required for a single binding/unbinding event considering different short simulation runs at 3000 K, 4000 K, 5000 K, 7000 K and 9000 K. This setup allowed the observation of multiple binding and unbinding events, while also ensuring adequate sampling along the CVs.  System's solvation, equilibration and production runs were performed following the same protocol and parameters described previously for the selection of the EN model. A total of 10 different replicas were performed, with 10 $\mu$s of sampling per replica, for a total simulation time of 100 $\mu$s. The final potential mean force (PMF) was obtained by integrating the mean forces on the auxiliary variables collected during TAMD runs, using the DRR plugin of PLUMED~\cite{Chen2018} and the Adaptive Biasing Force (ABF)~\cite{Darve2001,Lelivre2007,Lesage2016} \texttt{abf\_integrate} utility in Colvars~\cite{Fiorin2013}.

\subsection{Dimensionality reduction using 
$\beta$-Variational Autoencoder}

For dimensionality reduction, a binary contact map between CYFIP1 and eIF4E was computed for simulation frames sampled every 100 ps, yielding a total of $10^{6}$ structures. A contact was defined when the distance between the centers of mass of two residues was below 0.6 nm.
A set of distance matrices was then generated using only those residue 
pairs that showed more than 5000 contacts along the trajectory.
Frames with a value of the funnel-axis collective variable (CV) outside the interval between 0.1~nm and 4~nm were discarded.
Each distance matrix was flattened into a one dimensional feature vector, yielding a dataset of residue–residue distances comprising 850 features. The dataset was then standardized and used to train an autoencoder (AE), employed as a deterministic baseline, and a series of $\beta$-variational autoencoders ($\beta$VAE), each with a different $\beta$ value, employing the \texttt{mlcolvar} Python library \cite{Bonati2023}. 
The neural-network architecture consisted of an 
encoder with two hidden layers of 256 and 128 neurons, respectively, 
followed by a latent layer of two neurons. The decoder mirrored the 
encoder architecture. A shifted soft-plus activation function was used, 
with a learning rate of $5 \times 10^{-4}$, and a different $\beta$ factor for each model, ranging from $10^{-4}$ to $2$.

The AE was trained by minimizing the mean squared reconstruction
error over the standardized features,
\begin{equation}
  \mathcal{L}_{\mathrm{AE}}
  = \frac{1}{Nd}\sum_{i=1}^{N}
    \bigl\lVert \mathbf{x}_i-\hat{\mathbf{x}}_i \bigr\rVert^{2},
  \label{eq:ae_loss}
\end{equation}
where $\mathbf{x}_i \in \mathbb{R}^{d}$ is the $i$-th standardized
input vector ($d=850$), $N$ is the number of samples in the mini-batch,
$\hat{\mathbf{x}}_i = g_{\phi}(f_{\theta}(\mathbf{x}_i))$ is its
reconstruction, and $f_{\theta}$ and $g_{\phi}$ denote the encoder and
the decoder, respectively.

The $\beta$VAE models were trained by minimizing the negative
evidence lower bound with a Gaussian encoder and a standard normal
prior. The encoder outputs the mean $\boldsymbol{\mu}_i \equiv
\boldsymbol{\mu}(\mathbf{x}_i)$ and the log-variance
$\log\boldsymbol{\sigma}_i^{2}$ of the approximate posterior
$q_i \equiv q_{\theta}(\mathbf{z}\,|\,\mathbf{x}_i) =
\mathcal{N}(\boldsymbol{\mu}_i,\,
\mathrm{diag}\,\boldsymbol{\sigma}_i^{2})$, from which a latent point is
drawn through the reparameterization trick, $\mathbf{z}_i =
\boldsymbol{\mu}_i + \boldsymbol{\sigma}_i \odot
\boldsymbol{\epsilon}$ with $\boldsymbol{\epsilon} \sim
\mathcal{N}(\mathbf{0},\mathbf{I})$, and decoded into
$\hat{\mathbf{x}}_i = g_{\phi}(\mathbf{z}_i)$. The corresponding loss
function reads
\begin{equation}
  \begin{split}
    \mathcal{L}_{\beta\mathrm{VAE}}
    = {}& \frac{1}{Nd}\sum_{i=1}^{N}
          \bigl\lVert \mathbf{x}_i-\hat{\mathbf{x}}_i \bigr\rVert^{2} \\
      & + \beta\,\frac{1}{N}\sum_{i=1}^{N}
          D_{\mathrm{KL}}\bigl[\, q_i \,\big\Vert\, p(\mathbf{z}) \bigr],
  \end{split}
  \label{eq:vae_loss}
\end{equation}
where the first term accounts for the reconstruction accuracy and
the second one for the regularization of the latent space, with
$p(\mathbf{z})=\mathcal{N}(\mathbf{0},\mathbf{I})$. The
Kullback--Leibler divergence is evaluated analytically as
\begin{equation}
  \begin{split}
    D_{\mathrm{KL}}\bigl[\, q_i \,\big\Vert\, p(\mathbf{z}) \bigr]
    = -\frac{1}{2}\sum_{k=1}^{n_z} \Bigl(
      & 1 + \log\sigma_{i,k}^{2} \\
      & - \mu_{i,k}^{2} - \sigma_{i,k}^{2} \Bigr),
  \end{split}
  \label{eq:kl}
\end{equation}
where $n_z=2$ is the dimensionality of the latent space.

Intuitively, the encoder maps each high-dimensional input onto a compact, low-dimensional representation, but instead of collapsing it to 
a single fixed point, it maps the input onto a small region of the latent space, expressed as a probability distribution. A point is then 
randomly drawn from this region and passed to the decoder, which attempts to reconstruct the original input from it. This setup encourages two complementary behaviors. 
On one hand, the network is pushed to compress the 850 input features into just two latent variables while retaining enough structural information to allow an accurate reconstruction. On the other hand, it is also encouraged to keep the latent regions associated with different inputs close to a 
common, well-behaved reference distribution, rather than scattering them arbitrarily across the latent space. This second behaviour is what makes the latent space smooth and continuous, avoiding disconnected regions, and encourages the two latent dimensions to capture largely independent aspects of the input.
The parameter $\beta$ controls the balance between these two behaviours: 
larger values prioritize a cleaner, more structured latent space, at the 
cost of a less faithful reconstruction of the original features, while 
smaller values shift the model towards the behaviour of a standard, 
deterministic autoencoder, favoring reconstruction accuracy over 
regularization. By scanning a range of $\beta$ values, the most
informative latent representation of the CYFIP1--eIF4E dissociation was
selected. Training and validation loss curves for all models are reported in Fig.~S4, and the corresponding reconstruction accuracy, quantified as the error between input and reconstructed inter-residue distances, is reported in Fig.~S7.
After training, the best-performing model was used to predict the latent values for each
simulation frame.

The PMF surface in the latent variables, say $(z_1,z_2)$, was computed as
\begin{equation}
G(z_1,z_2) = - k_{\mathrm{B}} T \log P_{\mathrm{eq}}(z_1,z_2),
\end{equation} 
where \(P_{\mathrm{eq}}\) is the equilibrium probability density obtained by reweighting the TAMD trajectory:
\begin{equation}
P_{\mathrm{eq}}(\mathbf{z}) =
\frac{\left\langle \delta\!\left(\mathbf{z} - \boldsymbol{\xi}(\mathbf{x})\right)\, w(\mathbf{s}) \right\rangle_{\mathrm{TAMD}}}
{\left\langle w(\mathbf{s})\right\rangle_{\mathrm{TAMD}}} .
\end{equation}
Here, \(\boldsymbol{\xi}(\mathbf{x})=(\xi_1(\mathbf{x}),\xi_2(\mathbf{x}))\) denotes the CVs onto which the trajectory is projected (i.e. the latent variables), whereas \(\mathbf{z}=(z_1,z_2)\) denotes a point in the projected space. The brackets \(\langle \cdot \rangle_{\mathrm{TAMD}}\) indicate an average over the TAMD trajectory, and \(w(\mathbf{s})\) is the reweighting factor:
\begin{equation}
w(\mathbf{s}) = \exp\!\left[-(\beta - \bar{\beta})\, F(\mathbf{s})\right],
\end{equation}
where \(F(\mathbf{s})\) is the TAMD PMF along the original auxiliary variables \(\mathbf{s}\), while \(\beta=1/k_{\mathrm{B}}T\) and \(\bar{\beta}=1/k_{\mathrm{B}}\bar{T}\) are the inverse physical temperature and the effective inverse temperature of the auxiliary variables, respectively.

\subsection{Clustering of the latent space}
All points with free energy values greater than 1~kcal/mol on the reweighted FES were discarded, retaining only points within the various minima. Hierarchical clustering was then performed using Ward's method, with the linkage matrix computed from the latent space coordinates. The final number of clusters was selected based on the highest silhouette score.

The overall clustering performance is evaluated by computing two complementary measures of the average homogeneity of the resulting clusters, both derived from the normalized variance of interatomic distances within each cluster, as described below.

Let $S_1$ and $S_2$ be the two sets of particles (residues, represented as single beads in the coarse-grained model) corresponding to the most frequently contacted pairs identified in the contact analysis (see previous section). For a given cluster containing $N_f$ frames, we denote by $\mathbf{R}^{(f)}_i$ the position of particle $i$ at frame $f$. For each frame $f = 1, \dots, N_f$, we define the interatomic distance matrix:
\[
D^{(f)}_{ij} = \lVert \mathbf{R}^{(f)}_i - \mathbf{R}^{(f)}_j \rVert, 
\quad i \in S_1, \; j \in S_2 \; ,
\]
and the average distance matrix $\overline{D}_{ij}$ as:
\[
\overline{D}_{ij} = \frac{1}{N_f} \sum_{f=1}^{N_f} D^{(f)}_{ij} \; .
\]
The corresponding variance matrix is computed as:
\[
\sigma^2_{ij} = \frac{1}{N_f} \sum_{f=1}^{N_f} \left( D^{(f)}_{ij} - \overline{D}_{ij} \right)^2 \; ,
\]
from which we derive the normalized variance:
\[
\widetilde{\sigma^2_{ij}} = \frac{\sigma^2_{ij}}{\overline{D}_{ij}^2} \; .
\]

The $\widetilde{\sigma^2_{ij}}$ distributions for every cluster, obtained at the silhouette-optimal number of clusters for each reconstructed latent space (each $\beta$-VAE model, together with the AE and PCA cases), are reported in Figs.~S5 and S6, together with the cluster population and the silhouette profile as a function of the number of clusters.

The overall homogeneity is evaluated by computing the mean of the normalized variance across the different clusters:
% \begin{equation}
% \langle \widetilde{\sigma^2_{ij}} \rangle = \frac{1}{N_c} \sum_{c=1}^{N_c} \widetilde{\sigma^2_{ij}}^{(c)} \; ,
% \label{eq:mean_varianc
% \end{equation}

\begin{equation}
\langle \widetilde{\sigma^2} \rangle = 
\frac{1}{N_c} \sum_{c=1}^{N_c} 
\Big(
\frac{1}{N_{S_1}\times N_{S_2}}
\sum_{i,j} \widetilde{\sigma^2_{ij}}^{(c)}
\Big) \; ,
\label{eq:mean_variance}
\end{equation}

and the fraction of residue pairs with normalized variance below 0.1:
\begin{equation}
    \langle \theta \rangle = 
    \frac{1}{N_c} 
    \sum_{c=1}^{N_c}
    \Big(
    \frac{1}{N_{S_1}\times N_{S_2}}
    \sum_{i,j} \mathbf{1} 
    \left[ \widetilde{\sigma^2_{ij}}^{(c)} < 0.1 \right]\Big)
    \, ,
\label{eq:mean_fraction}
\end{equation}

with $N_{S_1}$ and $N_{S_2}$ the number of particles in the two selections. The metrics $\langle \widetilde{\sigma^2_{ij}} \rangle$ and $\langle \theta \rangle$ are reported in Fig.~\ref{fig:rew_pmf}b.

\section*{Data Availability Statement}
The data that support the findings of this study, including the PLUMED input files for the funnel-TAMD setup, representative structures for each cluster, and the training/analysis scripts for the dimensionality reduction models, will be openly available in Zenodo DOI:10.5281/zenodo.21809670 upon publication.

\section*{Author Contributions}
Francesco Pignotti: Conceptualization, Methodology, Software, Formal analysis, Investigation, Data curation, Writing -- original draft, Visualization. Emanuele Telari, Edoardo Milanetti: Validation, Writing -- review \& editing. Anna La Teana, Jesmina Rexha, Alice Romagnoli: Validation, Writing -- review \& editing. Luca Maragliano: Methodology, Supervision, Writing -- review \& editing. Giovanni Di Muccio: Conceptualization, Supervision, Funding acquisition, Project administration, Writing -- review \& editing.

\section*{Acknowledgments}
We acknowledge CINECA for the availability of high performance computing resources and support.

\section*{Declaration of Generative AI and AI-assisted
Technologies in the Writing Process}

During the preparation of this work the authors used a large
language model (Claude Opus 5, Anthropic) to proofread the
manuscript and assist in writing the analysis scripts. All generated text and
code were reviewed, verified and edited by the authors, who take full
responsibility for the scientific content of the publication.

\bibliographystyle{unsrt}
\bibliography{bibliography}
\end{document}